\documentclass[apj,numberedappendix]{emulateapj}
\usepackage{apjfonts}

\slugcomment{Accepted for publication in The Astrophysical Journal}
\shorttitle{THE ENERGY DEPENDENCE OF NEUTRON STAR SURFACE MODES}
\shortauthors{PIRO \& BILDSTEN}

\newcommand{\be}{\begin{eqnarray}}
\newcommand{\ee}{\end{eqnarray}}

\begin{document}

% -----------------------------------------------------------
% -----------------------------------------------------------

\title{The Energy Dependence of Neutron Star Surface Modes
and X-ray Burst Oscillations}

\author{Anthony L. Piro}
\affil{Department of Physics, Broida Hall, University of California
	\\ Santa Barbara, CA 93106; piro@physics.ucsb.edu}

\and

\author{Lars Bildsten}
\affil{Kavli Institute for Theoretical Physics and Department of Physics,
Kohn Hall, University of California
	\\ Santa Barbara, CA 93106; bildsten@kitp.ucsb.edu}

% -----------------------------------------------------------
% -----------------------------------------------------------

\begin{abstract}

  We calculate the photon energy dependence of the pulsed amplitude
of neutron star (NS) surface modes. Simple approximations demonstrate that
it depends most strongly on the bursting NS surface temperature. This
result compares well with full integrations that include Doppler shifts
from rotation and general relativistic corrections to photon propagation. 
We show that the energy dependence of type I X-ray burst oscillations 
agrees with that of a surface mode, lending further support to the
hypothesis that they originate from surface waves. The energy
dependence of the pulsed emission is rather insensitive to the NS
inclination, mass and radius, or type of mode, thus hindering
constraints on these parameters.  We also show that, for this
energy-amplitude relation, the majority of the signal (relative to the
noise) comes in the $\approx2-25\ {\rm keV}$ band, so that the
current burst oscillation searches with the {\it Rossi X-Ray Timing Explorer}
are close to optimal. The critical test of the mode hypothesis for X-ray
burst oscillations would be a measurement of the energy dependence
of burst oscillations from an accreting millisecond pulsar. 

\end{abstract}

\keywords{stars: neutron ---
	stars: oscillations ---
	X-rays: bursts ---
	X-rays: stars}

% -----------------------------------------------------------
% -----------------------------------------------------------

\section{Introduction}

  Accreting neutron stars (NSs) often show nearly coherent
modulations during type I X-ray bursts \citep[see reviews of][]{bil98,sb03}
called burst oscillations \citep[][and references therein]{mun01}.
The high temporal stability of each NS's characteristic frequency
\citep[within 1 part in $10^3$ over years,][]{mun02}, along with burst
oscillations seen from two accreting millisecond pulsars at their spin
frequencies \citep{cha03,str03} have led many to conclude that burst
oscillations are a modulation at the NS spin frequency. Nevertheless,
it has been a long standing mystery as to what creates the surface
asymmetry in the burst tail, long after any hot spots from the burst
ignition should have spread over the entire star \citep{bil95,slu02}.

  A recently developed and promising hypothesis is that burst
oscillations are surface {\it r}-modes \citep{hey04}. In this picture
the oscillations are created by a retograde mode with an observed
frequency just below the NS spin. As the star cools in the burst
tail the mode replicates the observed rising frequency. Current
theoretical work has focused on calculating the frequencies
\citep{lee04,pb05b} and flux perturbations \citep{hey05,ls05} expected
for such modes. Unfortunately, besides the highly sinusoidal nature
of burst oscillations \citep{moc02}, which is expected for modes, there
is little direct evidence that modes are the correct explanation.
\citet{pb05b} addressed this issue by calculating how the mode's
properties would manifest themselves in the burst oscillations. First, they
showed that higher persistent luminosity NSs should exhibit smaller
frequency drifts, consistent with current observations. Second, they
hypothesized that additional modes might be present
with such large frequency drifts that they are difficult to detect.
Though these are promising steps, both predictions are directly tied
to Piro \& Bildsten's (2005b) model that the burst oscillations are a
surface wave changing into a crustal interface wave \citep{pb05a}.
What is needed is a complementary, and more general, argument
of how a surface mode should exhibit itself, independent of the specific
model invoked. That is the goal of this present study.

  A key characteristic of burst oscillations is a larger pulsed fraction
at increasing energies in the range of $2-23\ {\rm keV}$
\citep[][hereafter MOC03]{moc03}. This is distinct from other pulsing
NSs. For example, accretion-powered millisecond X-ray pulsars
decrease (marginally) in amplitude between $2-10\ {\rm keV}$
\citep{cmt98,gal02,pg03}. MOC03 modeled the burst oscillations
as a hot spot on the NS surface and nicely  reproduced the
burst oscillation energy dependence. Given the promising possibility
that the oscillation may be a mode, we perform a calculation of the
pulsed amplitude as a function of energy for a nonradial oscillation.
Similar studies have recently been completed by \citet{hey05} and
\citet{ls05}. Our work differs from these in that we are primarily
interested in comparisons with observations, and thus do not perform
an exhaustive parameter survey. We also include analytic estimates
to explain our numerical calculations.

  A primary difficulty with comparisons between theory and observations
is that the mode amplitudes are not predicted from linear mode calculations.
To overcome this complication we follow the example of pulsating white
dwarf (WD) studies \citep{kep00} and only fit the {\it shape} of the energy
dependence, leaving the overall amplitude as a free parameter. Unfortunately,
bursting NSs have a drawback with respect to pulsating WDs in that
the limb darkening of their bursting envelope is largely independent of
photon energy above $\approx1\ {\rm keV}$ \citep{mad91}. 
This prevents constraining
NS properties with burst oscillation data because the inclination,
NS mass, $M$,  and radius, $R$,  and even the mode's angular eigenfunction only
alter the pulsed amplitude normalization as we show in
\S \ref{sec:comparison}. For this reason, full integrations of the pulsed
NS emission are well replicated by an analytic result
that only depends on the NS surface color
temperature, $T_c$ (eq. [\ref{eq:result}]).

  We compare our result with the observed energy dependence of burst
oscillations, finding agreement for $k_{\rm B}T_c\approx2-3\ {\rm keV}$, as
expected for NSs during X-ray bursts. This suggests that
the burst oscillations are due to a nonradial mode, independent of the mode's
identification. The excitation and nonlinear evolution
of the mode is of upmost importance if we are to infer NS attributes from
burst oscillations.

 In \S \ref{sec:theory} we calculate the energy dependence
of a mode's amplitude and compare it with X-ray
burst oscillations in \S \ref{sec:observations}. In \S \ref{sec:sn} we investigate
the optimal  photon energy ranges for detection of burst oscillations. 
We conclude in \S \ref{sec:conclusion} 
with a summary of our results and note the importance of 
measuring the energy dependence of burst oscillations from an 
accreting millisecond pulsar.

% -----------------------------------------------------------
% -----------------------------------------------------------

\section{The Energy Dependence of a Mode's Amplitude}
\label{sec:theory}

  We first describe our procedure for calculating the pulsed amplitude
of a surface mode as a function of energy. This 
follows previous studies of pulsed NS emission \citep[for example,][]{pg03},
but is included to provide context for our analytic results in
\S \ref{sec:analytic}. In \S \ref{sec:comparison}
we compare the analytics with numerical integrations.

% -----------------------------------------------------------

\subsection{Equations for Calculating the Pulsed Amplitude}
\label{sec:full}

  We use a spherical coordinate system given by $(r,\theta,\phi)$ for
the inertial reference frame of the observer, with its origin at the center
of the star. The observer sits at an angle $\theta=0$. Nonradial
oscillations are set in a spherical coordinate system that shares
its origin with the observer's coordinates, but is rotated by an inclination
angle, $i$. We denote this by $(r,\theta',\phi')$, with the pulsation axis,
which is coincident with the spin axis, at $\theta'=0$. The cartesian
coordinates of the two frames are related by
\newcounter{subequation}[equation]
\renewcommand{\theequation}{\arabic{equation}\alph{subequation}}
\be
	\addtocounter{subequation}{+1}
	x' &=& x\cos i - z\sin i
	\\
	\addtocounter{equation}{-1}
        \addtocounter{subequation}{+2}
	y' &=& y
	\\
	\addtocounter{equation}{-1}
        \addtocounter{subequation}{+3}
	z' &=& x\sin i + z\cos i.
\ee
Gravitational light bending causes photons that reach the observer
to be emitted at an angle $\alpha\geq\theta$, which is given 
to high accuracy for a Schwarzschild metric by
\be
	1-\cos\alpha = (1-\cos\theta)\left(1-\frac{r_g}{R}\right),
\ee
where $r_g=2GM/c^2$ is the Schwarzschild radius \citep{bel02}. 

   The equations that we derive adopt a number of simplifying
 assumptions.
These are justified because they either are negligible corrections in the context
of pulsed emission from a NS, or because they do not affect the energy
dependence of the pulsed amplitude. We ignore
the effects of frame dragging, which would modify our results by an amount
less than the current observational errors (Cadeau et al. 2005 show that a
Schwarzschild plus Doppler treatment as presented here provides sufficiently
accurate results in comparison to a full relativistic calculation). Lorentz
boosting can be ignored since it is a $\lesssim0.1\%$ correction for
a NS spin of
$\nu=600\ {\rm Hz}$, and as well we omit relativistic aberration because
it only marginally alters our results.
Finally, we also ignore Doppler shifting from the wave motions because the
transverse velocity of {\it r}-modes \citep[$\sim10^7\ {\rm cm\ s^{-1}}$ for an
order unity perturbation, approximated from the results of][]{pb05b}
is much less than $c$. 

  Given these simplifications the observed flux at photon energy $E$ is
related to the intensity from the surface, $I(E',\theta',\phi')$, by \citep{pg03}
\be
	F(E) \propto
	\int\int
	\delta^3
	I(E',\theta',\phi')h(E',\cos\alpha)
		\cos\alpha d\Omega,
	\label{eq:integral}
\ee
where $h(E',\cos\alpha)$ is the limb darkening function,
$d\Omega=d\cos\alpha d\phi$ is the angular element,
and $E'=E/\left(\delta\sqrt{1-r_g/R}\right)$ is the photon energy 
in a frame co-rotating with the NS surface.
The Doppler factor is given by
\be 
        \delta = 1/(1-\beta\sin\alpha\sin\phi\sin i),
        \label{eq:dopplerfactor}
\ee
where $\beta=2\pi R\nu/\left(c\sqrt{1-r_g/R}\right)$ is the equatorial velocity (with $\nu$ the spin
frequency). In addition, equation (\ref{eq:integral}) should have
factors due to gravitational redshift, the NS radius, and the NS
distance, but we omit these since they cancel when we take the
pulsed fraction. The integration limits are $0\leq\alpha\leq\pi/2$ and
$0\leq\phi\leq2\pi$. 

  In general, the limb darkening function can depend on photon energy.
This is crucial for studies of ZZ Ceti stars, which have opacities strongly
affected by lines, so that the latitudinal quantum numbers can be identified
by studying the energy dependence of the pulsed emission
\citep{rkn82,kep00}. In contrast, bursting NSs have a surface
opacity dominated by electron
scattering. For photons with energy $\gtrsim1\ {\rm keV}$ the limb darkening is
largely energy independent and well-approximated by $h=0.5+0.5\cos\alpha$
\citep{mad91}, which is the functional form we assume for our
calculations.

  Finding the perturbed flux requires perturbing each term in 
the integrand of equation (\ref{eq:integral}) and keeping terms
of linear order. This results in three integrals to evaluate, which
correspond to changes in intensity, surface area, and
surface normal. Since for nonradial incompressible modes the transverse
velocity dominates over the radial velocity ($V_\perp/V_r\sim R/H\gg1$, where
$H$ is the scale height in the bursting layer), the latter two changes are
negligible \citep{bs79,rkn82}. Using just the integral which
contains the intensity perturbation, $\Delta I$,
the fractional amplitude of the mode is then
\be
        A(E)&\equiv&\frac{\Delta F(E)}{F(E)}
        \nonumber
        \\
	&=&\frac{\displaystyle \int\int\delta^3
		\Delta I(E',\theta',\phi') h(\cos\alpha)\cos\alpha d\Omega}{\displaystyle
		\int\int\delta^3
                I(E',\theta',\phi') h(\cos\alpha)\cos\alpha d\Omega}.
        \label{eq:ampl1}
\ee
We next relate $\Delta I$ to the mode eigenfunction, which
is just the temperature perturbation. This relation depends
on the bursting NS spectrum, which is well-characterized as a dilute blackbody with a
temperature given by a color temperature $T_c\approx(1.4-1.6)T_{\rm eff}$
\citep{mad91,psz91,mad97}. The change in overall normalization does not
affect the energy dependence, hence 
we use $I(E')=B_{E'}(T_c)$ and perturb
this by setting  $T_c\rightarrow T_c+\Delta T$, keeping terms of
linear order in $\Delta T$,
\be
	\frac{\Delta I}{I} = \frac{\partial\log I}{\partial\log T_c}
		\frac{\Delta T}{T_c}
		=
		\frac{x' e^{x'}}{e^{x'}-1}
		\frac{\Delta T}{T_c},
	\label{eq:intensity}
\ee
where $x'\equiv E'/k_{\rm B}T_c$.
Substituting this result into equation
(\ref{eq:ampl1}), the fractional amplitude becomes
\be
	A(E) =
		\frac{\displaystyle \int\int \frac{x' e^{x'}}{e^{x'}-1}
		\frac{\Delta T}{T_c}(\theta',\phi')I(E')
		h(\cos\alpha)\cos\alpha d\Omega}{\displaystyle
			\int\int I(E')h(\cos\alpha)
			\cos\alpha d\Omega},
	\label{eq:ampl2}
\ee
which can be integrated for  any angular eigenfunction
$\Delta T(\theta',\phi')/T_c$.

% -----------------------------------------------------------

\subsection{Analytic Estimates}
\label{sec:analytic}

  Before we calculate equation (\ref{eq:ampl2}) numerically we simplify
the integrals so that their energy dependence can be studied analytically.
There only exists an energy dependence in two terms: $I(E')$ and the
logarithmic derivative found in equation (\ref{eq:intensity}). If these terms
contained no angular dependence, they could be taken outside of the integrals
so that the integrals become irrelevant for determining the energy
dependence of $A(E)$. In principle
this cannot be done because $E'$ contains an angular dependence through
the Doppler factor, equation (\ref{eq:dopplerfactor}), so that 
the integrals must be performed numerically.

  On the other hand, if the Doppler shifts are negligible,
then the energy dependence of $A(E)$ is simply
\be
	A(E)	\propto \frac{x'e^{x'}}{e^{x'}-1}.
	\label{eq:result}
\ee
This has high and low energy limits that are useful
for gaining intuition about the expected dependence on energy,
\be
       A(E) \propto \left\{
              \begin{array}{cc}
        E/k_{\rm B}T_c,
                &\hspace{0.3cm}E>k_{\rm B}T_c\sqrt{1-r_g/R} \\
        {\rm constant},
                &\hspace{0.3cm}E<k_{\rm B}T_c\sqrt{1-r_g/R}
              \end{array}
       \right.
       \label{eq:limits}
\ee
At high energies the amplitude is linear with
energy, while at low energies the amplitude is approximately
constant. This argument shows why a mode will naturally show a larger
amplitude at larger energy. {\it It is simply a result of perturbing a blackbody
spectrum, and the fractional change of intensity is much stronger in the
Wien tail.} Similar results are found by MOC03 for their hot spot model
when the temperature difference between hot and cold regions is small.
When the temperature difference they use becomes large, this can lead to
deviations away from a linear amplitude-energy relation at high energies
($x'\gtrsim10$), perhaps
providing an important discriminant between the mode and hot spot
models. Unfortunately, this is outside the currently observed energy range.

  To find the correction to equation (\ref{eq:result}) introduced by Doppler
effects we expand $\Delta I$ to first
order in $\beta'\equiv\beta\sin\alpha\sin\phi\sin i$, giving
\be
	\Delta I \approx (\Delta I)_{\beta=0}
	\left[ 1+\beta'\left(\frac{2x'e^{x'}}{e^{x'}-1}-x'-1\right)\right].
	\label{eq:doppler}
\ee
When $x'\ll1$, this changes the amplitude by a factor $1+\beta'$, which
contains no energy dependence. When $x'\gg1$ the
amplitude changes by a factor $1+\beta'x'$, so that Doppler shifting 
increases the amplitude at larger energy. 

% -----------------------------------------------------------

\subsection{Comparisons to the Numerical Integrations}
\label{sec:comparison}

  We now compare our analytic results to numerical
integrations of equation (\ref{eq:ampl2}). This shows that the analytics
reproduce the amplitude versus energy relation. For the angular pattern of the mode,
$\Delta T(\theta',\phi')/T_c$, we use a buoyant
{\it r}-mode with angular quantum numbers $l=2$ and $m=1$ \citep[as
identified in the slowly rotating limit,][]{pb04}, a favored
mode for reproducing the frequency evolution of the burst oscillations
\citep{hey04, hey05, pb05b}. This mode's latitudinal eigenfunction is
parametrized by the spin parameter $q=2\nu/f$, where $f$ is the mode
frequency in a frame co-rotating with the star in units of ${\rm Hz}$. A faster spinning
NS (higher $q$) results in an eigenfunction that is more concentrated
near the NS equator due to Coriolis effects. We fix
$M=1.4\ M_\odot$ and $k_{\rm B}T_c=3\ {\rm keV}$ so that
we can concentrate on whether other attributes of a NS can affect the energy
dependence. The amplitude at a given energy is assessed by calculating
the time-dependent amplitude and then fitting this with a sinusoidal function.

\begin{figure}
\epsscale{0.95}
%\epsscale{0.8}
\plotone{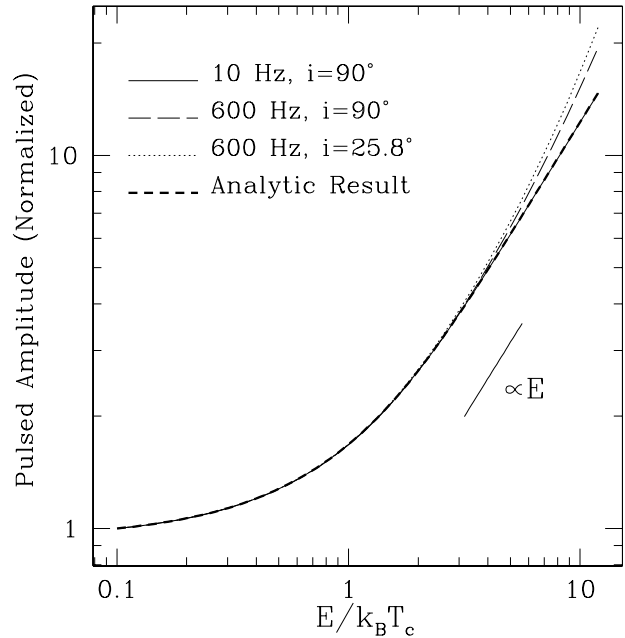}
\caption{The energy dependence of the pulsed amplitude, $A(E)$, for both the full
numerical integration and the analytic result given by equation (\ref{eq:result})
({\it thick dashed line}).
The numerical integrations all use a $q=200$, $l=2$, $m=1$ buoyant
{\it r}-mode on a $M=1.4\ M_\odot$ and $R=10\ {\rm km}$ NS.
The parameters we explore are $\nu=10\ {\rm Hz}$, $i=90\degr$ ({\it solid line}),
$\nu=600\ {\rm Hz}$, $i=90\degr$ ({\it long dashed line}), and $\nu=600\ {\rm Hz}$,
$i=25.8\degr$ ({\it dotted line}).
Though the normalization can change drastically for different inclinations
\citep[for an example, see Fig. 4 of][]{hey05} we
renormalize all the results to $A(E)=1$ at $E/k_{\rm B}T_c=0.1$ to
focus on the shape of the energy dependence. At high energies, $A(E)\propto E$
as we show in \S \ref{sec:analytic}.}
\label{fig:spin}
\epsscale{1.0}
\end{figure}

  In Figure \ref{fig:spin} we compare integrations of different spins and inclinations,
keeping the mode pattern fixed at $q=200$ as well as fixing $M$ and $R$. 
All amplitudes are normalized to $A(E)=1$ at $E/k_{\rm B}T_c=0.1$.
At low spin, $\nu=10\ {\rm Hz}$ ({\it solid line}), the analytic
result of equation (\ref{eq:result}) ({\it thick dashed line}) and the numerical calculation
are practically identical. As the spin is increased to $\nu=600\ {\rm Hz}$
({\it long dashed line}) the amplitude increase at high energies, as predicted by equation (\ref{eq:doppler}). We also consider a more
face-on orientation for the NS ($i=25.8\degr$, {\it dotted line}), which has the effect of
looking like a faster spin NS. This somewhat counterintuitive result has been
seen in previous studies \citep{hey05} and is due the
mode pattern, which has a maximum amplitude at latitudes
above and below the equator. These comparisons
show that neither the spin nor inclination change the energy dependence
dramatically from our analytic result.

\begin{figure}
\epsscale{0.95}
%\epsscale{0.8}
\plotone{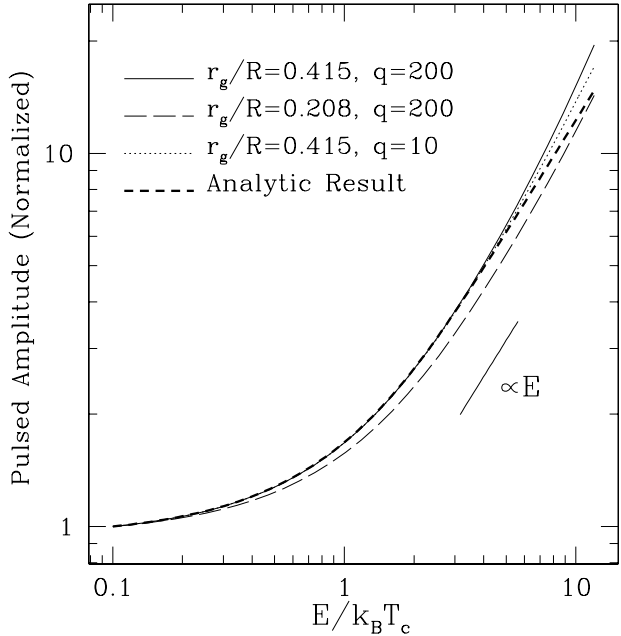}
\caption{The same as Figure \ref{fig:spin}, but with $\nu=600\ {\rm Hz}$ and $i=90\degr$
and varying $q$ and $R$ for the numerical calculations. Setting $R=20\ {\rm km}$
({\it long dashed line}) has the effect of decreasing the amplitude
at high energies (compare this to the solid line).
Decreasing $q$ from 200 to 10 ({\it dotted line}) affects the energy dependence very little, showing
that it is difficult to constrain the surface pattern created by the mode.}
\label{fig:nsproperties}
\epsscale{1.0}
\end{figure}

  In Figure \ref{fig:nsproperties} we keep the spin and inclination fixed at
$\nu=600\ {\rm Hz}$ and $i=90\degr$, and investigate the effect of changing $q$ and
$R$. When we set $R=20\ {\rm km}$ ({\it long dashed line}) the
amplitude of the pulsed fraction decreases at high
energies. This is because changing $R$ decreases gravitational redshifting
so that the break between a constant and linearly increasing amplitude comes
at a higher energy (see eq. [\ref{eq:limits}]). We also decrease
$q$ dramatically ({\it dotted line}) but find very little change in the energy dependence.
This shows that it is difficult to identify the mode pattern on the NS surface via
the amplitude energy dependence. 

% -----------------------------------------------------------
% -----------------------------------------------------------

\section{Comparisons with Observations}
\label{sec:observations}

  MOC03 studied the energy dependence of burst oscillation amplitudes
from 6 different bursting NSs. A total of 51 burst oscillation trains were measured,
and multiple trains were averaged to obtain amplitude
versus energy relations.  Some objects were divided into
multiple epochs to assure that the gain of the Proportional
Counter Array (PCA) on the {\it Rossi X-Ray Timing Experiment} ({\it RXTE}) was
relatively constant. To correctly compare our calculations to their results
we must weight our pulsed amplitudes by the PCA's effective area,
$A_{\rm eff}(E)$, as well as bin the amplitudes over appropriate energy ranges.
The PCA is composed of five Proportional Counter Units
(PCUs). Since each of these have approximately the same $A_{\rm eff}(E)$,
we use that of PCU3 for our integrations.
The pulsed amplitude is then given by equation (\ref{eq:ampl2}) with
$A_{\rm eff}(E)$ placed within each integrand. Qualitatively, $A_{\rm eff}(E)$
has a large wide maximum spread from $\approx4-15\ {\rm keV}$ with a smaller,
secondary peak at $\approx34\ {\rm keV}$. The binning of the amplitude
depends on the epoch of the observation and is outlined in Table 2 of MOC03.

\begin{figure*}
%\epsscale{1.1}
\epsscale{1.2}
\plotone{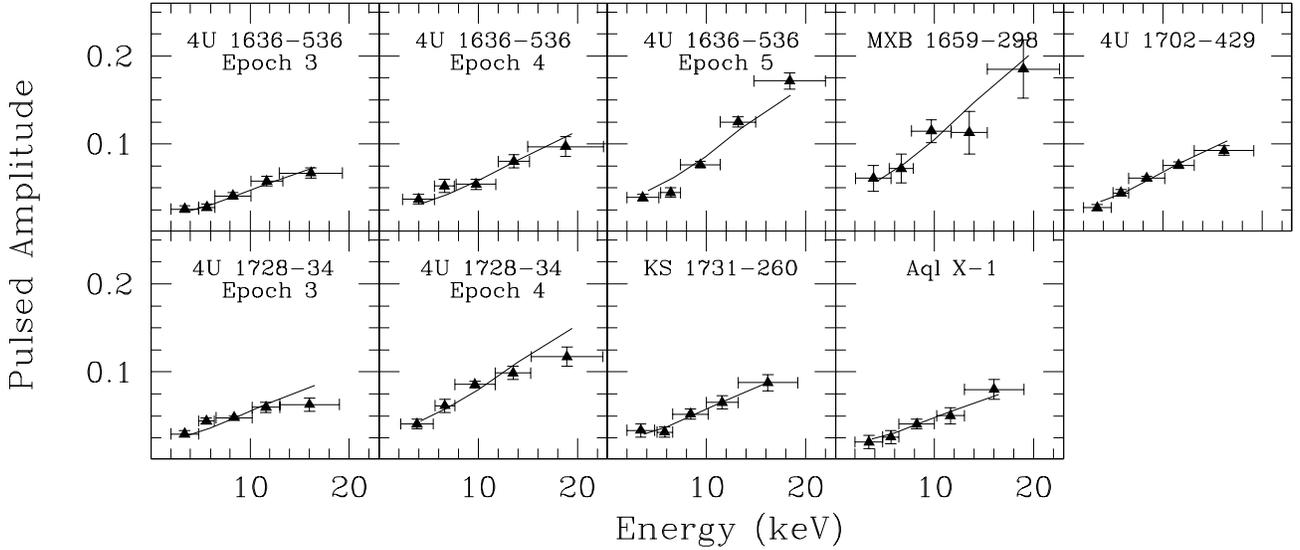}
\caption{The observed energy dependence of the burst oscillation amplitudes
(MOC03, {\it triangles with error bars}),
in comparison with our full numerical calculations using
$k_{\rm B}T_c=3\ {\rm keV}$ ({\it solid lines}).
We label each panel with the NSs name 
and the epoch of the measurements if that NS was observed over multiple epochs.
%The observed amplitudes within each panel have
%been normalized to unity at the highest energy bin of that panel.
For each calculation the spin is set to that
object's burst oscillation frequency and the normalization is set to maximize
the fit.}
\label{fig}
\end{figure*}

    In Figure \ref{fig} we compare the calculated amplitudes with the
measurements of MOC03 ({\it triangles with error bars}). For the
calculated amplitudes we fix $M=1.4\ M_\odot$, $R=10\ {\rm km}$, $q=200$, and
$i=90$. The spin is set to the burst oscillation frequency for that
object. This is reasonable since in all current mode explanations of burst
oscillations the mode moves retrograde with respect to the spin with
$\nu\gg f$ \citep[see discussion in][]{ls05}.
%In each panel the observed pulsed amplitude is normalized to unity
%at the highest energy bin.
The normalization of the numerical calculations
are set to maximize the fit for each comparison. We consider
$k_{\rm B}T_c=3\ {\rm keV}$ ({\it solid lines}) as a fiducial temperature
exhibited near burst peak. Ideally, we
should be able to constrain $k_{\rm B}T_c\sqrt{1-r_g/R}$ by fitting for
the break in the
amplitude (providing $M/R$ if $T_c$ is known). This is difficult because
when the photon energy is
$\gtrsim k_{\rm B}T_c$, as is the case for these observations, the
amplitude is always linear with energy. Nevertheless, the theoretical calculations show reasonable agreement with the observations.

\begin{figure}
\epsscale{0.95}
%\epsscale{0.8}
\plotone{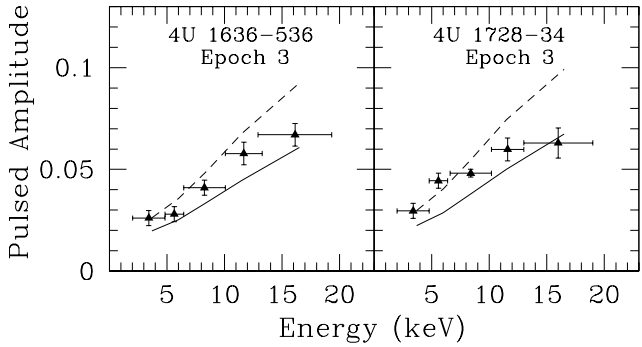}
\caption{The observed energy dependence of the burst oscillations amplitudes
(MOC03, {\it triangles with error bars}), for 2  of the 9 panels
from Fig. \ref{fig}. These represent a relatively ``good'' fit
({\it left panel}) and a ``poor'' fit ({\it right panel}).
 We compare these with our numerical calculations using
$k_{\rm B}T_c=2\ {\rm keV}$ ({\it dashed lines}) and $3\ {\rm keV}$ ({\it solid lines}).
The overall normalization is set to maximize the fit with the data,
but the relative amplitude of the two curves is set by the two temperatures. This
shows that the observed spread in that data could arise 
from the cooling of the NS in the
burst tail.}
\label{fig:2}
\epsscale{1.0}
\end{figure}

  Comparisons to the observations are complicated by the fact that the observations
are averaged over a range of temperatures throughout the cooling of the burst, so
that we should consider temperatures in the range of $k_{\rm B}T_c\approx2-3\ {\rm keV}$.
If the mode amplitude remains relatively constant, then the
pulsed amplitude in the Wien tail should increase as the star cools (see eq.[\ref{eq:limits}]).
We test this in Figure \ref{fig:2} for two of the observed amplitudes from Figure \ref{fig},
but in this case calculating the amplitudes for temperatures of
$k_{\rm B}T_c=2\ {\rm keV}$ ({\it dashed lines})
and $3\ {\rm keV}$ ({\it solid lines}). The overall normalization is chosen to maximize
the fit, but the relative amplitude of the two curves in each panel
is set by the temperature ratio. The two temperatures envelop the data, showing that
the spread in the data may be due to cooling in the burst tail. An interesting future
test of our work would be to divide the burst data into early and late stages,
and see whether the amplitude evolves. This expected temperature
dependence can be divided out of the data to investigate how much the amplitude of the
mode is changing due to other effects (e.g., changes in the intrinsic amplitude,
or changes in $q$).

% -----------------------------------------------------------
% -----------------------------------------------------------

\section{Optimal Energies for Detection of Neutron Star Modes}
\label{sec:sn}

\begin{figure}
%\epsscale{0.9}
\epsscale{1.0}
\plotone{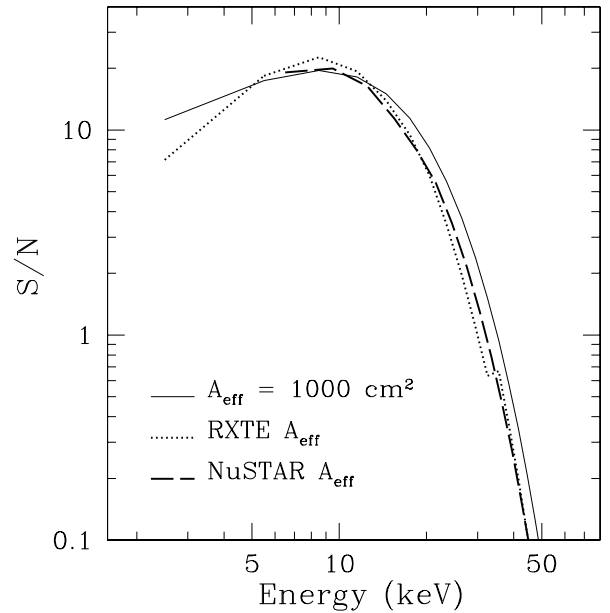}
\caption{The ratio of signal to noise (using eqs. [\ref{eq:signal}] and [\ref{eq:noise}])
expected for a NS surface mode during a $10\ {\rm s}$ X-ray burst. We set
$k_{\rm B}T_c=3\ {\rm keV}$, $D=4.3\ {\rm kpc}$, and $\Delta T/T_c=0.025$,
and compare a flat energy response
({\it solid line}), the $A_{\rm eff}(E)$ of one PCU on {\it RXTE}'s PCA
({\it dotted line}), and the proposed $A_{\rm eff}$ for the {\it NuSTAR} mission ({\it dashed line}).
Each curve connects a series of points, with each point representing
an integration over a $3\ {\rm keV}$ width energy bin.
}
\label{fig:sn}
\epsscale{1.0}
\end{figure}

  Since we understand the spectrum of the burst
oscillations, this can be used to find the optimal photon
energy range for burst oscillations searches. The pulsed signal is given by
the total number of pulsed photons integrated over some energy range,
\be
	S &=&\sqrt{1-\frac{r_g}{R}}\left(\frac{R}{D}\right)^2t_{\rm obs}\int A_{\rm eff}(E)\frac{dE}{E}
		\nonumber
		\\
	&&\times\int\int\frac{x' e^{x'}}{e^{x'}-1}
		\frac{\Delta T}{T_c}I(E')
		h(\cos\alpha)\cos\alpha d\Omega,
	\label{eq:signal}
\ee
where $D$ is the source distance, $t_{\rm obs}$ is the observing time,
and we have assumed $\delta\approx1$.
The background noise within this energy range is estimated from
photon counting statistics 
\be
	N & = & \left[\sqrt{1-\frac{r_g}{R}}\left(\frac{R}{D}\right)^2t_{\rm obs}
		\int A_{\rm eff}(E)\frac{dE}{E}\right.
		\nonumber
		\\
		&&\left.\times\int\int I(E')
		h(\cos\alpha)\cos\alpha d\Omega\right]^{1/2},
	\label{eq:noise}
\ee
which is the square-root of the total number
of photons detected,

  To evaluate equations (\ref{eq:signal}) and (\ref{eq:noise}) we use
a blackbody spectrum, $I(E') = B_{E'}(T_c)$, with $k_{\rm B}T_c=3\ {\rm keV}$,
and the analytic pulsed fraction, ignoring
Doppler corrections, since the agreement is so close between the numerical and analytic
results. We assume $D=4.3\ {\rm kpc}$
(using 4U $1728-30$ for a fiducial distance), $t_{\rm obs}=10\ {\rm s}$
(one X-ray burst) and $\Delta T/T_c=0.025$ (which replicates the observed
pulsed fractions). 
In Figure \ref{fig:sn} we plot these S/N calculations using three different forms for $A_{\rm eff}$:
a flat energy response with $A_{\rm eff}=1000\ {\rm cm^{2}}$
({\it solid line}), the $A_{\rm eff}$ of one PCU from {\it RXTE}'s PCA ({\it dotted line}),
and the proposed $A_{\rm eff}$ for the
{\it Nuclear Spectroscopic Telescope Array} \citep[{\it NuSTAR};][{\it dashed line}]{har04},
a future X-ray mission.
For each we show a series of $3\ {\rm keV}$ integrations, which are then connected
with lines to guide the eye.
This comparison demonstrates that the
energy range of $\approx2-25\ {\rm keV}$ contributes most to $S/N$.
Other than integrating over this range,
there is little more an observer can do to maximize the
opportunity of finding burst oscillations. At high energies the pulsed
fraction is considerably higher, but this range is not necessarily
better since there are so few photons in the
Wien tail and because $A_{\rm eff}$ drops at higher energies.
Future missions
interested in burst oscillations searches could mitigate this
by having large $A_{\rm eff}$ at higher energies. 

  It is especially exciting that {\it NuSTAR} may be able to observe burst
oscillations. Though {\it NuSTAR}'s specifications of a large $A_{\rm eff}$,
high energy range ($\approx5-80\ {\rm keV}$), and good spectral
resolution (900 eV at 68 keV) are tuned for
observations of black holes, active galactic nuclei, and supernova
remnants, it also has fast timing ($\sim1\ {\rm ms}$) which makes
it ideal for studying burst oscillations. In addition, its high angular resolution
($\approx40\ {\rm arcsec}$) coupled with its timing abilities may make it
useful for identifying accreting millisecond pulsars in crowded fields such as at
the Galactic center (something beyond {\it RXTE}'s capabilities).
A typical accreting millisecond pulsar at the galactic center has a peak flux in
outburst 100 times less than a type I burst, so the persistent
pulse could easily be found in a $\approx1\ {\rm day}$ long observation.

% -----------------------------------------------------------
% -----------------------------------------------------------

\section{Discussion and Conclusions}
\label{sec:conclusion}

  We have studied the energy dependence of NS surface mode amplitudes
for NS surface temperatures of $k_{\rm B}T_c=3\ {\rm keV}$
and compared this with burst oscillations. The observations follow our
calculated trend of a linear amplitude for photon energies
$\gtrsim k_{\rm B}T_c$ and becoming flatter when the photon energy is
$\approx k_{\rm B}T_c$.
Unfortunately, there are currently no data for burst oscillations below an energy
of $k_{\rm B}T_c$. Measuring the amplitude at such an energy is crucial for
fully testing the mode explanation of burst oscillations. One must be cautious
of interpretations for $E'\lesssim1\ {\rm keV}$ because limb darkening
begins to depend on energy in this range, so that equation (\ref{eq:result})
is no longer applicable.  However, this also raises the 
possibility for surface pattern identification 
at low energies
\citep[analogous to what is done for pulsating WDs,][]{kep00}.

  The agreement between our analytic result and the observations is promising
for explaining burst oscillations as modes, but frustrates the ability of using burst
oscillations as a tool to learn about these NSs. As long as the amplitude of the surface
perturbation caused by the mode is unknown, it will be difficult to constrain NS
properties. Future theoretical studies
should work to address this unanswered question. We expect all bursting NS modes
to show the energy dependence we present here, including the oscillation seen
in the 4U $1636-53$ superburst \citep{sm02}, for which this energy dependence
was not determined.

  However, other types of NS pulsations need not match this, for example millisecond
accreting pulsars in their persistent emission. This raises the critical question
of what is the energy dependence of the burst oscillations from these systems
(SAX J$1808.4-3658$ [Chakrabarty et al. 2003] and
XTE J$1814-338$ [Strohmayer et al. 2003]).
\citet{pb05b} describe a number of differences
between the properties of burst oscillations from pulsars and nonpulsars.
These differences may simply be due to deviations in magnetic field strength, but
they could instead indicate that the pulsar burst oscillations are due to a completely
different mechanism. Measuring their energy dependence would help to settle
this unresolved issue.

  One topic we have not addressed is the phase lag observed for high energy
photons in burst oscillations (MOC03). We focused on the amplitude relation
because of its stronger statistical evidence in the observations. 
For individual oscillation trains measured by MOC03, only 13 out of 51 exhibit
phases that vary as a function of energy at 90\% confidence. In comparison,
34 of these exhibit some dependence with energy at 90\% confidence (with the
remaining typically having less counts in their folded profile). 
Nevertheless, it is somewhat troubling that our calculations find the reverse
phase dependence due to Doppler shifts, just as was found in previous studies of
pulsed emission from NSs \citep[e.g.,][]{wml01,hey05}.
This inconsistency has been cited by many as evidence that a Comptonizing
corona exists around a NS during an X-ray burst. Without further theoretical
studies of the physical limits of such a Comptonizing corona or further investigations
on how robust of a property this observed hard phase lag is, it is not presently clear how
dire this inconsistency is for the mode explanation of burst oscillations.
\citet{ls05} have
found some parameter space in $R$ and $i$ that exhibit hard lags, which
may be promising to pursue further. Though
their result is for a different eigenfunction than we consider here, it does not affect
our main conclusions since the energy dependence will still be independent of the
specific mode.
%In fact, this energy dependence was also found from our calculations of
%the pulsed emission expected from the shearing modes studied by \citet{cum05}.
%We were interested in seeing whether this mode has the desired hard phase
%lag, but this was not exhibited by the low-order even mode that \citet{cum05}
%found to be unstable.

\acknowledgements
  We appreciate the help of Mike Muno for providing us with the pulsed amplitude data
and Philip Chang for reading and contributing comments on a previous draft.
We thank Fiona Harrison for providing us with the effective area response of
the {\it NuSTAR} satellite.
We also thank Deepto Chakrabarty and Andrew Cumming for many helpful
discussions. This work was supported by the
National Science Foundation under grants PHY99-07949 and AST02-05956,
and by the Joint Institute for Nuclear Astrophysics through NSF grant PHY02-16783.

% -----------------------------------------------------------
% -----------------------------------------------------------

\end{document}